# Source Enumeration via RMT Estimator Based on Linear Shrinkage Estimation of Noise Eigenvalues Using Relatively Few Samples

Huiyue Yi, *Member, IEEE*

*Abstract*—Estimating the number of signals embedded in noise is a fundamental problem in array signal processing. The classic RMT estimator based on random matrix theory (RMT) tends to under-estimate the number of signals as it does not consider the non-negligible bias term among eigenvalues for finite sample size. Moreover, as the eigenvalue being tested is assumed arising from a signal in the RMT estimator, the noise variance will be under-estimated when the eigenvalue being tested is actually arising from noise. Therefore, the RMT estimator will over-estimate the number of signals, and thus suffers from uncertainty in noise variance estimation. problem. In order to overcome these problems, we firstly derive a more accurate expression for the distribution of the sample eigenvalues and the bias term among eigenvalues by utilizing the linear shrinkage (LS) estimate of noise sample eigenvalues. Then, we analyze the effect of the bias term among eigenvalues on the estimation performance of the RMT estimator, and derive the increased under-estimation probability of the RMT estimator incurred by this bias term. Based on these results, we propose a novel RMT estimator based on LS estimate of noise eigenvalues (termed as "LS-RMT estimator") by incorporating the bias term into the decision criterion of the RMT estimator. As the LS-RMT estimator incorporates this bias term among eigenvalues into the decision criterion of the RMT estimator, it can detect signal eigenvalues immersed in this bias term. Therefore, the LS-RMT estimator can overcome the higher under-estimation probability of the RMT estimator incurred by the bias term among eigenvalues. Moreover, the LS-RMT can also avoid the uncertainty in the noise variance estimation suffered by the RMT estimator as the noise variance is estimated under the assumption that the eigenvalue being tested is arising from noise. Finally, extensive simulation results are presented to show that the proposed LS-RMT estimator outperforms the existing estimators.

## I. Introduction

ESTIMATING the number of signals in a linear mixture model is a fundamental problem in statistical signal processing and array signal processing [1]-[6]. The classical approaches for signal number estimation are derived based on information theoretic criteria (ITC) [7], including AIC [8], MDL [9], etc. A more detailed review of ITC-based approaches can be found in [10]. Basically, these methods rely on the asymptotic distribution of the sample eigenvalues known from the classical multivariate statistical theory, and have been developed for the classical asymptotic regime in which the system size $p$ is fixed while the number of samples $n$ tends to infinity. It is implied that the ratio $p/n \to 0$. In practice, the dimensionality of the data in modern signal processing applications is very large and the acquisition of more observations is often costly. Therefore, the general asymptotic regime, where $p, n \to \infty$ with $p/n \to c \in (0, \infty)$, is more suitable to large-array applications when large antenna arrays are employed or when the sample size and the system size are finite with comparable values [11]-[13]. For example, the massive MIMO utilizes large antenna arrays with up to a few hundred antennas to improve spectral efficiency [12]-[13], and has become a key technique for 5G wireless communication systems. These large-scale arrays typically operate in the general asymptotic scenario where the number of antennas is comparable with the number of samples. In the general asymptotic scenario, however, the sample covariance matrix has a quite large variance and thereby is an inaccurate estimate of the covariance matrix and the corresponding eigenvalues. Therefore, the ITC-based methods may suffer significant performance degradation, and may be unable to estimate the source number accurately and efficiently.

In [14]-[15], the authors derived expressions for the extreme values of the perturbed eigenvalues (or singular values) of finite, low rank perturbations of large random matrices. It is shown that the perturbed eigenvalues (or singular values) tend to be increased relative to the true/unperturbed values if and only if the true eigenvalues are above a certain critical threshold. This phenomenon motivates a shrinkage-based approach, i.e., applying an adaptive shrinkage to the eigenvalues/singular values to efficiently improve the estimates of the covariance matrix of the signal and noise subspace components by shrinking the sample covariance [16]-[18]. In [17], the authors present a RBLW estimator for shrinkage of noise eigenvalues in the sense of minimum mean-squared error (MMSE). In [18], the authors derived the optimal shrinkage coefficients for the signal eigenvalues. Therefore, the shrinkage approach can be utilized to enable the ITC-based approaches to properly detect the source number in the general asymptotic regime. In [19], the authors utilize the linear shrinkage (LS) technique to obtain

Manuscript submitted on October 27, 2020. This work was supported by the Shanghai Municipal Key project under Grant 19511132401, and Shanghai Municipal Natural Science Foundation under Grant 18ZR1437600.

Huiyue Yi is with the Shanghai Institute of Microsystem and information technology, Chinese Academy of Science, 3/F, Xinwei Building A, 1455 Pingcheng Road, Jiading, Shanghai 201800, P. R. China (e-mail: huiyue.yi@mail.sim.ac.cn; huiyue_yi@263.net).

an accurate estimator for the noise covariance matrix, and propose a LS-MDL estimator for signal number estimation. In [20], the authors estimate the latent low-rank signal matrix by hard thresholding the singular values (SVHT) of the noisy data matrix by employing Stein's unbiased risk estimator (SURE) [21], and then the number of SVs above the threshold can be seen as an estimate of the signal number. In [22], an effective rank selection method is proposed by employing SURE. As is illustrated in [23], the SURE-based rank estimators with SVHT and SVST result in a sub-optimal estimate of the rank. In SVHT, noise in signal dominated components is neglected and this may cause an error in the rank selection. When the SNR is low, the SVST is unable to shrink the noise dominated SVs to zero and may wrongly over-estimate the noise SVs as signal SVs. In [24], the authors have derived the optimal nonlinearities of the shrinkage of the SVs in the recovery of low-rank matrices from noisy data.

The large dimensional random matrix theory [25]-[26] has emerged as a powerful tool to deal with the general asymptotic scenario when classical multivariate statistical theory fails. In fact, the large random matrix theory has been widely applied in the research of massive MIMO systems [27]-[29]. In massive MIMO systems, the noise or signal subspace cannot be correctly determined for the general asymptotic scenario. To cope with this problem, more efficient subspace-based algorithms [30]-[31] have been proposed for the large array. Moreover, estimating the number of signals embedded in noise is the prerequisite step for these subspace-based algorithms in massive MIMO systems. However, most conventional source enumeration techniques cannot properly work and provide poor results in the general asymptotic scenario. Therefore, the source number estimation method should be further studied in massive MIMO or large-scale array systems. The random matrix theory concerns both the distribution of noise eigenvalues and of the signal eigenvalues in the large-system-size large-sample-size asymptotic region [32]-[42]. As is justified by these works, though the random matrix theory is based on the high-dimension and large-sample asymptotic regime, it provides a more accurate approximation for the distribution of the sample eigenvalues for the practical scenarios where the sample size and the system size are finite and probably comparable in magnitude than the classical multivariate statistical theory. In recent years, the use of random matrix theory in estimating the number of signals or weak signal detection the general asymptotic scenario has attracted much attention [43]-[50]. In these methods, results on the spectral behavior of random matrices are applied to the problem of detecting the number of signals in a noisy data matrix. As shown in [32]-[35], the fluctuation of the largest noise eigenvalue of the sample covariance matrix can be modeled by the Tracy-Widom distribution. Based on this result, the authors in [44] proposed a RMT estimator to estimate the number of signals via sequentially detecting the largest noise eigenvalues as arising from a signal or noise for a given over-estimation probability. In [45], the authors utilized the joint distribution of the sum of the noise eigenvalues and sum of the squares of noise eigenvalues and developed the RMT-AIC criterion for signal number estimation in the general asymptotic case. However, the RMT-AIC estimator cannot provide the consistent estimate of the source number [44]. In [46], the authors proposed a sample generalized eigenvalue-based detection of signals (termed as GE-based estimator hereafter) using relatively few signal-bearing and noise-only samples. In [47], the authors analyze the detection performance of the AIC estimator from the random matrix theory point, and propose a modified AIC estimator with a small increase in the penalty term. However, finding the optimal penalty term for the AIC and MDL is still an open question to be solved. In [48], the authors propose a signal number estimator based on the difference between two consecutive sample eigenvalues (denoted as EV-Difference estimator), and sequentially test the eigenvalue as arising from a signal or from noise via checking whether this difference is above a pre-fixed threshold or not. In [49], the authors take into account the distributions of the noise and signal eigenvalues, and derive a two-step test method for signal number estimation. However, the likelihood test in its second step is a sub-optimal test for finite sample sizes since it neglects the dependence between the eigenvalues. In [50], the authors propose an improved MDL estimator based on higher-order moments of the sample eigenvalues (termed as HO-MDL estimator).

As proved in [50]-[51], there exists a non-negligible bias term among eigenvalues when the number of samples is relatively small. In [52], the authors provide an insightful analysis on the bias term on the missed detection probability of the MDL estimator by utilizing the results from the random matrix theory. However, the aforementioned methods, including shrinkage-based methods and RMT-based methods, do not consider the bias term among eigenvalues. Therefore, the weak signal eigenvalues will be immersed in the largest noise eigenvalues, and thus the number of signals tends to be under-estimated. To our best knowledge, the impact of this bias term among eigenvalues on the detection performance of the existing signal number estimators is still an open problem to be solved. Moreover, the noise variance cannot be known a priori and instead needs to be estimated. Therefore, the uncertainty in the noise variance estimation inevitably degrades the performance of these methods.

In this paper, we will focus on developing an effective signal number estimation method that is free of the uncertainty in the noise variance estimation and can detect the signal eigenvalues immersed in the bias term among eigenvalues for finite sample size. To achieve this aim, in this paper we derive a more accurate expression for the distribution of the sample eigenvalues and the bias term among eigenvalues by utilizing the linear shrinkage estimator (LS) in [19], and then propose an RMT estimator based on the LS estimate of the sample eigenvalues (termed as "LS-RMT estimator") by incorporating the bias term among eigenvalues into the decision criterion of the RMT estimator. In the development of the LS-RMT estimator, we utilize the results regarding the asymptotically norm distribution of the sample signal eigenvalues [38]-[41]. In addition, we utilize the results regarding the expectations of the sample signal eigenvalues [50]-[51], which reflects the interaction among eigenvalues for limited sample size and system size. As the proposed LS-RMT estimator incorporates the bias term into the decision criterion, it can successfully detect the signal eigenvalues immersed in the bias term among eigenvalues. Moreover, the LS-RMT estimator also overcomes the uncertainty in noise variance estimation suffered by the RMT estimator. The main contributions of this work are summarized as follows:

(1) As illustrated in [17]-[19], the sample eigenvalue corresponding to noise has a large variance and is usually ill conditioned in the general asymptotic regime. In order to overcome this problem, In Section III. A we derive a more accurate expression for the distribution of the sample eigenvalues by utilizing the LS estimator in [19]. Specifically, we derive the analytical formulas for both the bias term among eigenvalues and the variance of the sample eigenvalues. Then, in Section III.B we analyze the effect of this bias term among eigenvalues on the estimation performance of the RMT estimator. Specifically, we derive the analytical formulas for the increased under-estimation probability of the RMT estimator incurred by the bias term among eigenvalues.

(2) In Section III. C, we propose a novel LS-RMT estimator by incorporating the bias term among eigenvalues into the decision criterion of the RMT estimator. Firstly, we calculate the increased under-estimation probability of the RMT estimator incurred by the bias term among eigenvalues. Based on these results, the proposed LS-RMT estimator incorporates the bias term among eigenvalues into the decision criterion of the RMT estimator. Therefore, the LS-RMT estimator can successfully detect signal eigenvalues immersed in this bias term, and thus avoids the higher under-estimation probability of the RMT estimator. In addition, the noise variance is estimated under the assumption that the eigenvalue being tested is arising from noise in the LS-RMT estimator, and thus the LS-RMT estimator can also overcome the uncertainty in the noise variance estimation suffered by the RMT estimator.

(3) Extensive simulations are presented to compare the estimation performance of the proposed LS-RMT estimator to the existing methods including the AIC and MDL in [7]-[9], the LS-MDL estimator in [19], the RMT estimator [44], and the modified AIC estimator [47], the GE-based estimator in [46], the EV-Differ estimator in [48], and the HO-MDL estimator in [50]. Simulation results show that the proposed LS-RMT estimator, the proposed LS-RMT estimator outperforms the existing estimators.

This paper is organized as follows. In Section II, we present the problem formulation, the mathematical preliminaries from the random matrix theory and the prior works. In Section III, we firstly derive a more accurate expression for the distribution of the sample eigenvalues and the bias term among eigenvalues by utilizing the linear shrinkage estimator in [19], and analyze the effect of the bias term among eigenvalues on the estimation performance of the RMT estimator. Then, we describe the proposed LS-RMT estimator. Simulation results that illustrate the superior estimation performance of the proposed LS-RMT estimator over the existing methods are presented in Section IV. Finally, conclusions are drawn in Section V.

*Notation*: In this paper, we use boldface uppercase letters to denote matrices, boldface lowercase letters for column vectors, and lowercase letters for scalar quantities. Superscripts $(\cdot)^T$ and $(\cdot)^H$ represent transpose and conjugate transpose, respectively. Furthermore, $\hat{a}$ denotes the estimate of $a$, $\mathbf{I}_p$ is the $p \times p$ identity matrix, $\mathbf{0}_p$ is the $p \times 1$ zero vector, and $\mathbf{x} \sim \mathcal{N}(\boldsymbol{\mu}, \boldsymbol{\Sigma})$ ($\mathbf{x} \sim \mathcal{CN}(\boldsymbol{\mu}, \boldsymbol{\Sigma})$) denotes that $\mathbf{x}$ follows a real (complex) Gaussian distribution with mean $\boldsymbol{\mu}$ and covariance matrix $\boldsymbol{\Sigma}$. The notations $\xrightarrow{D}$ and $\xrightarrow{a.s.}$ denote convergence in distribution, and almost surely, respectively, the symbol $\mathrm{E}[\cdot]$ denotes expectation, and the function $\mathrm{Pr}(\cdot)$ denotes probability.

## II. PROBLEM FORMULATION, RANDOM MATRIX THEORY AND PRIOR WORKS

In this section, we firstly introduce the data model and problem formulation. Then, we provide the mathematical preliminaries from the random matrix theory and the Lawley's law. Finally, we describe the RMT estimator in [44].

### A. Data Model and Problem Formulation

We consider the standard model $p$-dimensional linear mixture model for signals impinging on an array with $p$ sensors [44]. Let $\{\mathbf{x}(i) = \mathbf{x}(t_i)\}_{i=1}^n$ denote $n$ independent and identically distributed (i.i.d.) observations of the form

$$\mathbf{x}(t) = \sum_{i=1}^q \mathbf{v}_i s_i(t) + \mathbf{w}(t) = \mathbf{A}\mathbf{s}(t) + \mathbf{w}(t), \quad (1)$$

where, $\mathbf{s}(t) = [s_1(t), \cdots, s_q(t)]^T$ is a $q \times 1$ vector containing $q$ different zero-mean signal components with corresponding independent array response vectors $\mathbf{v}_i \in \mathbb{R}^p$, $\mathbf{A} = [\mathbf{v}_1, \mathbf{v}_2, \cdots, \mathbf{v}_q]$, and the noise $\mathbf{w}(t)$ are assumed to be additive white Gaussian noise (AWGN) with zero mean and unknown variance $\sigma^2$, i.e., $\mathbf{w}(t) \sim N(\mathbf{0}_p, \sigma^2 \mathbf{I}_p)$, and $\mathbf{w}(t)$ is independent of $\mathbf{s}(t)$. Under these assumptions, the population covariance matrix of the observations $\mathbf{x}$ is given by $\boldsymbol{\Sigma} = E[\mathbf{x}\mathbf{x}^H]$ with its $q$ noise-free population signal eigenvalues given by $\{\lambda_1, \lambda_2, \cdots, \lambda_q\}$, and thus the population eigenvalues $\{\rho_j\}_{j=1}^p$ of $\boldsymbol{\Sigma}$ is given by

$$\{\lambda_1 + \sigma^2, \cdots, \lambda_q + \sigma^2, \sigma^2, \cdots, \sigma^2\}. \quad (2)$$

In practice, the population covariance matrix $\boldsymbol{\Sigma}$ is estimated using only a finite number of samples, namely, the sample covariance matrix. Given $n$ i.i.d. samples $\{\mathbf{x}(i)\}_{i=1}^n$, the sample covariance matrix $\mathbf{S}_n$ is given by

$$\mathbf{S}_n = \frac{1}{n}\sum_{i=1}^{n}\mathbf{x}(i)\mathbf{x}^H(i). \tag{3}$$

Let the sample eigenvalues of $\mathbf{S}_n$ be $l_1 \geq l_2 \geq \cdots \geq l_p$. Then, the problem is to determine the number $q$ of signals from finite noisy samples $\{\mathbf{x}(i)\}_{i=1}^n$. In the general asymptotic scenario, i.e., $p, n \to \infty$ with $p/n \to \gamma \in (0, \infty)$, the RMT-based methods [5]-[6], [43]-[50] utilize the results from the random matrix theory, and have superior estimation performance over the classical ITC-based methods. However, these methods either suffer from noise uncertainty problem, either do not consider the inter-action terms among eigenvalues. To overcome these problems, in this paper we will further consider inferring the unknown number $q$ of signals from the $n$ samples $\{\mathbf{x}(i)\}_{i=1}^n$ under the nonparametric setting from the viewpoint of the random matrix theory.

*B. Mathematical Preliminaries: Random Matrix Theory and Lawley's Law*

In most cases, the number of sources is much smaller than the system size, i.e., $q \ll p$, which means that the population covariance matrix $\Sigma = E[\mathbf{xx}^H]$ is a low rank perturbation of an identity matrix. Such a population covariance matrix is called as the spiked covariance model [36]-[42], where all eigenvalues of the population covariance matrix are equal except for a small fixed number of distinct "spike eigenvalues". As the key goal in nonparametric estimation of the number of sources is to distinguish between noise and signal eigenvalues, in this subsection we will review some related results under this spiked covariance model regarding the asymptotic distribution of the eigenvalues of the sample covariance matrix $\mathbf{S}_n$.

The first result describes the asymptotic distribution of the largest eigenvalue of a pure noise matrix [32]-[35]. Let $\mathbf{S}_n$ denote the sample covariance matrix of pure noise observations distributed as $N(\mathbf{0}_p, \sigma^2\mathbf{I}_p)$. In the joint limit $p, n \to \infty$ with $p/n \to \gamma \in (0, \infty)$, the distribution of the largest eigenvalue of $\mathbf{S}_n$ converges to the Tracy-Widom distribution. That is, for every $x \in \mathbb{R}$,

$$\Pr[l_1 < \sigma^2(\mu_{n,p} + x\sigma_{n,p})] \to F_\beta(x). \tag{4}$$

where $\beta = 1$ for real valued noise and $\beta = 2$ for complex-valued noise. The centering and scaling parameters $\mu_{n,p}$ and $\sigma_{n,p}$, respectively, are functions of $n$ and $p$ only [32]-[35]. For real valued noise, the following formulas provide $O(p^{-2/3})$ convergence rate in (4), see [34]

$$\mu_{n,p} = \frac{1}{n}(\sqrt{n-1/2} + \sqrt{p-1/2})^2, \tag{5}$$

$$\sigma_{n,p} = \sqrt{\frac{\mu_{n,p}}{n}}\left(\frac{1}{\sqrt{n-1/2}} + \frac{1}{\sqrt{p-1/2}}\right)^{1/3}. \tag{6}$$

The second result characterizes the limiting distributions of the signal eigenvalues with strength $\lambda_j > \sigma^2\sqrt{\gamma}$ [14]-[15] and [38]-[41]. Such signal eigenvalues are distributed normally around the limiting value $(\lambda_j + \sigma^2)(1 + \gamma\sigma^2/\lambda_j)$. For the $j$th signal with $\lambda_j > \sigma^2\sqrt{\gamma}$ ($j = 1, 2, \cdots, q$), in the general asymptotic regime, where $p, n \to \infty$ with $p/n \to \gamma \in (0, \infty)$, at a convergence rate of $O(n^{-1/2})$, the fluctuations of $l_j$ converges with probability one to the normal distribution [44]:

$$l_j^{\text{asy}} \xrightarrow{D} N(\tau(\lambda_j), \delta^2(\lambda_j)), \tag{7}$$

where the superscript "asy" indicates $l_j^{\text{asy}}$ is a asymptotic expression for $l_j$, and $\xrightarrow{D}$ denotes convergence in distribution, and

$$\tau(\lambda_j) = (\lambda_j + \sigma^2)\left(1 + \frac{p-q}{n} \cdot \frac{\sigma^2}{\lambda_j}\right), \tag{8}$$

$$\delta(\lambda_j) = (\lambda_j + \sigma^2)\sqrt{\frac{2}{\beta n}\left(1 - \frac{p-q}{n} \cdot \frac{\sigma^4}{\lambda_j^2}\right)}. \tag{9}$$

While asymptotically there are no signal-signal interactions among signal eigenvalues, we need to take into account the non-negligible interaction among eigenvalues for finite $p$ and $n$. In [50]-[51], a more accurate expression for the expectation value of the sample eigenvalue $l_j$ for $j \leq q$ in the non-asymptotic region is given by:

$$E[l_j] = \rho_j + \frac{(p-q)\rho_j\sigma^2}{n(\rho_j - \sigma^2)} + \frac{\rho_j}{n}\sum_{i=1, i\neq j}^{q}\frac{\rho_i}{\rho_j - \rho_i} + O(n^{-2}) \tag{10}$$

where $\rho_j = \lambda_j + \sigma^2$. As can be seen from (10), the sample eigenvalues are affected by a bias term for finite $n$, which is caused by the interaction among the eigenvalues. For the sake of simplification, we let

$$\nu_j = \frac{\rho_j}{n}\sum_{i=1, i\neq j}^{q}\frac{\rho_i}{\rho_j - \rho_i}, \tag{11}$$

$$\kappa_j = 1 + \frac{(p-q)\sigma^2}{n\lambda_j}. \tag{12}$$

Then, (10) can be rewritten as

$$E[l_j^{\text{n-asy}}] = \rho_j \cdot \kappa_j + \nu_j + \mathcal{O}(n^{-2}). \quad (13)$$

*C. Prior Works*

As stated in (4), the fluctuation of the largest noise eigenvalue of the sample covariance matrix can be modeled by the Tracy-Widom distribution under the assumption of Gaussian data. Consequently, if the noise variance $\sigma^2$ is known, a statistical procedure to distinguish a signal eigenvalue $l$ from noise at a significant level $\alpha$ is to check whether $l > \sigma^2(\mu_{n,p} + s(\alpha)\sigma_{n,p})$, where the value of $s(\alpha)$ depends on the required significant level $\alpha$. Here, the significance level $\alpha$ is the probability of rejecting a null hypothesis when it is true. Based on this observation, a RMT estimator was proposed in [44] to estimate the number of signals via detecting the largest noise eigenvalues.

The RMT estimator is based on a sequence of hypothesis tests, for $k = 1, 2, \cdots, \min(p,n) - 1$,

$$\begin{aligned} H_k &: \text{at least } k \text{ signals,} \\ H_{k-1} &: \text{at most } k-1 \text{ signals.} \end{aligned} \quad (14)$$

For each value of $k$, the noise level $\sigma^2$ and the signal eigenvalue $\{\lambda_i\}_{i=1}^k$ are estimated assuming that $l_{k+1}, \cdots, l_p$ correspond to noise via solutions of the following non-linear system of equations [44]:

$$\sigma_{\text{RMT}}^2(k) - \frac{1}{p-k}\left[\sum_{j=k+1}^{p} l_j + \sum_{j=1}^{k}(l_j - \hat{\rho}_j)\right] = 0, \quad (15)$$

$$\hat{\rho}_j^2 - \hat{\rho}_j\left[l_j + (1 - \frac{p-k}{n})\sigma_{\text{RMT}}^2(k)\right] + l_j\sigma_{\text{RMT}}^2(k) = 0. \quad (16)$$

After the convergence of the above system, we obtain the estimates for $\{\hat{\rho}_i\}_{i=1}^k$ and noise level $\sigma_{\text{RMT}}^2(k)$, and the signal eigenvalue $\lambda_i$ is estimated as $\hat{\lambda}_i = \hat{\rho}_i - \sigma_{\text{RMT}}^2(k)$. Then, the likelihood of the $k$th eigenvalue $l_k$ is tested as arising from a signal or from noise as follows:

$$l_k > \sigma_{\text{RMT}}^2(k)\left(\mu_{n,p-k} + s(\alpha)\sigma_{n,p-k}\right). \quad (17)$$

For this test to have a false alarm with asymptotic probability $\alpha$ as $p, n \to \infty$, the threshold $s(\alpha)$ should satisfy

$$F_\beta(s(\alpha)) = 1 - \alpha. \quad (18)$$

The value of $s(\alpha)$ can be calculated by inverting the Tracy-Widom distribution, and this inversion can be computed numerically by using the software package [53]. If (17) is satisfied, $H_k$ is accepted and $k$ is increased by one. Otherwise, $\hat{q} = k - 1$. That is to say

$$\hat{q} = \arg\min_k \left\{ l_k < \sigma_{\text{RMT}}^2(k)\left(\mu_{n,p-k} + s(\alpha)\sigma_{n,p-k}\right)\right\} - 1. \quad (19)$$

As can be seen from (13), there exists a non-negligible bias term among eigenvalues when the number of sample size $n$ is limited. However, the decision criterion in (17) of the RMT estimator does not consider this bias term among eigenvalues, and thus the estimation performance of the RMT estimator will be affected by this bias term for finite $p$ and $n$. As will be analyzed in Section III. B, the RMT estimator tends to have higher under-estimation probability as some signal eigenvalues will be immersed in this bias term. Moreover, the noise variance $\sigma_{\text{RMT}}^2(k)$ in (17) is estimated under the assumption that $l_k$ is arising from a signal, and thus will be under-estimated when $l_k$ is actually arising from noise. In turn, the under-estimated noise level $\sigma_{\text{RMT}}^2(k)$ will lead to higher over-estimation probability of the RMT estimator. Therefore, the RMT estimator suffers from the uncertainty in noise variance estimation. In order to overcome these problems, in Section III we will develop an LS-RMT estimator by utilizing the linear shrinkage estimator of noise eigenvalues in [19], the results regarding the asymptotically norm distribution of the sample signal eigenvalues given by (7) and the expectation value of the sample eigenvalues given by (13) for finite $p$ and $n$.

## III. RMT Estimator via Linear Shrinkage Estimator of Noise Eigenvalues for Signal Number Estimation

In this Section, we firstly derive a more accurate expression for the distribution of the sample eigenvalues and the bias term among eigenvalues by utilizing the linear shrinkage estimate of noise eigenvalues in [19]. Then, we analyze the effect of the non-negligible bias term among eigenvalues on the estimation performance of the RMT estimator [44] for finite $p$ and $n$. Based on these analysis results, we propose an RMT estimator based on LS estimate of noise eigenvalues (termed as "LS-RMT estimator") by incorporating this bias term into the RMT estimator. Finally, we provide performance comparison of the proposed LS-RMT estimator with the RMT estimator.

*A. Derivation for the distribution of sample eigenvalues via linear shrinkage estimator*

From (13), the mean of $l_q$ is given by

$$E[l_q] = \rho_q \cdot \kappa_q + \nu_q + \mathcal{O}(n^{-2}). \quad (20)$$

From (8) and (12), the mean of $l_q^{\text{asy}}$ is given by

$$E[l_q^{asy}] = \tau(\lambda_q) = \rho_q \cdot \kappa_q. \tag{21}$$

On the other hand, in the joint limit $p, n \to \infty$, with $p/n \to \gamma \in (0, \infty)$, from (9) the variance of $l_q^{asy}$ is given by

$$\text{var}[l_q^{asy}] = \delta^2(\lambda_q). \tag{22}$$

From (20) and (21), we can express $E[l_q]$ as

$$E[l_q] = E[l_q^{asy}] + v_q + \mathcal{O}(n^{-2}). \tag{23}$$

As illustrated in [17]-[19], the sample eigenvalue $l_q$ corresponding to noise has a large variance and is usually ill conditioned in the general asymptotic regime, In the following, we will assume $\rho_q = \sigma^2$ (i.e., $\lambda_q = 0$) and derive a more accurate expression for $l_q$ and the bias term $v_q + \mathcal{O}(n^{-2})$ by utilizing the linear shrinkage technique in [19].

From [19] (Eq. (29)), an accurate estimate for the eigenvalue $l_q$ by utilizing the linear shrinkage (LS) estimator is given by

$$\rho_q^{LS} = \beta_{LS}^{(q-1)} \hat{\tau}^{(q-1)} + (1 - \beta_{LS}^{(q-1)}) l_q, \tag{24}$$

where $\hat{\tau}^{(q-1)} = 1/(p-q+1) \cdot \sum_{i=q}^{p} l_i$, $\beta_{LS}^{(q-1)} = \min(\hat{\alpha}_{LS}^{(q-1)}, 1)$. The optimal shrinkage coefficient $\hat{\alpha}_{LS}^{(q-1)}$ is given by [19] (Eq. (27)):

$$\hat{\alpha}_{LS}^{(q-1)} = \frac{\sum_{i=q}^{p} l_i^2 + (\sum_{i=q}^{p} l_i)^2}{(n+1)\left(\sum_{i=q}^{p} l_i^2 - \frac{(\sum_{i=q}^{p} l_i)^2}{p-(q-1)}\right)}. \tag{25}$$

From (24), we have $\rho_q^{LS} = \hat{\tau}^{(q-1)}$ when $\beta_{LS}^{(q-1)} = 1$. Therefore, we define an indicator function $\mathbf{1}_{(\beta_{LS}^{(q-1)} < 1)}$ to indicate whether the shrinkage is applied to $l_q$. Then, (24) can be rewritten as

$$\rho_q^{LS} = \hat{\tau}^{(q-1)} - (1 - \beta_{LS}^{(q-1)})(\hat{\tau}^{(q-1)} - l_q) \cdot \mathbf{1}_{(\beta_{LS}^{(q-1)} < 1)}. \tag{26}$$

Then, we let

$$\Delta \rho_q = \rho_q^{LS} - \rho_q$$
$$= (\hat{\tau}^{(q-1)} - \sigma^2) - (1 - \beta_{LS}^{(q-1)})(\hat{\tau}^{(q-1)} - l_q) \cdot \mathbf{1}_{(\beta_{LS}^{(q-1)} < 1)}. \tag{27}$$

Obviously, the real value of the noise eigenvalue $l_q$ will be increased by a value of $\Delta \rho_q$. Therefore, a more accurate expression for $l_q$ is given by

$$l_q = l_q^{asy} + v_q + \Delta \rho_q = l_q^{asy} + v_q + (\hat{\tau}^{(q-1)} - \sigma^2) - (1 - \beta_{LS}^{(q-1)})(\hat{\tau}^{(q-1)} - l_q) \cdot \mathbf{1}_{(\beta_{LS}^{(q-1)} < 1)}. \tag{28}$$

Then, $l_q$ can be expressed as

$$l_q = \frac{l_q^{asy} + v_q - \sigma^2}{1 - (1 - \beta_{LS}^{(q-1)}) \cdot \mathbf{1}_{(\beta_{LS}^{(q-1)} < 1)}} + \hat{\tau}^{(q-1)}$$
$$= [1 + (\frac{1}{\beta_{LS}^{(q-1)}} - 1) \cdot \mathbf{1}_{(\beta_{LS}^{(q-1)} < 1)}](l_q + v_q - \sigma^2) + \hat{\tau}^{(q-1)}. \tag{29}$$

It follows from (25) that $1/\hat{\alpha}_{LS}^{(k)} - 1$ can be expressed as

$$\frac{1}{\hat{\alpha}_{LS}^{(k)}} - 1 = \frac{(n+1)(z-1)}{z+(m-k)} - 1 \triangleq h(z), \tag{30}$$

where

$$z = \left[\frac{1}{p-k} \sum_{i=k+1}^{p} l_i^2\right] / \left[\frac{1}{p-k} \sum_{i=k+1}^{p} l_i\right]^2. \tag{31}$$

It is shown in [45] that $z - z_0 \xrightarrow{D} \mathcal{N}(0, 2\gamma^2/(p-k)^2)$ with $z_0 = 1 + \gamma$.

Using the Taylor series expansion of $h(z)$ around $z_0$, we obtain

$$h(z) = h(z_0) + h'(z_0)(z - z_0) + \frac{1}{2} h''(z_0)(z - z_0)^2 + \cdots, \tag{32}$$

where $h'(z_0)$ and $h''(z_0)$ are the first-order and second-order derivatives of $h(z)$. Moreover, the parameter $\gamma$ can be replaced by $(p-k+1)/(n+1)$ for large $p$ and $n$ [19].. Therefore, we can set $\gamma = (p-k+1)/(n+1)$. Then, it is easy to derive the following equations:

$$h(z_0) = \frac{(n+1)\gamma}{\gamma+1+m-k} - 1 = \frac{-1}{n+2}, \tag{33}$$

$$h'(z_0) = \frac{(n+1)^2}{\gamma(n+2)^2}, \tag{34}$$

$$h''(z_0) = -2\frac{(n+1)^2}{\gamma^2[n+2]^3}. \tag{35}$$

Then, $h(z)$ in (32) is calculated as

$$h(z) = \frac{-1}{n+2} + \frac{(n+1)^2}{\gamma[n+2]^2}(z - z_0) - \frac{(n+1)^2}{\gamma^2[n+2]^3}(z - z_0)^2 + \cdots. \tag{36}$$

From (36), we can derive the following equations:

$$E[(1+(\frac{1}{\beta_{LS}^{(q-1)}}-1)\cdot \mathbf{1}_{(\beta_{LS}^{(q-1)}<1)})]$$

$$= E[(1+(\frac{1}{\alpha_{LS}^{(q-1)}}-1)\cdot \mathbf{1}_{(\alpha_{LS}^{(q-1)}<1)})] \quad , \quad (37)$$

$$= 1+[\frac{-1}{n+2}-\frac{2(n+1)^2}{(n+2)^3(p-q+1)^2}]\cdot \mathbf{1}_{(\beta_{LS}^{(k)}<1)}$$

$$\triangleq P(n,p,q)$$

$$E[(1+(\frac{1}{\beta_{LS}^{(q-1)}}-1)\cdot \mathbf{1}_{(\beta_{LS}^{(q-1)}<1)})^2]$$

$$= E[(1+(\frac{1}{\alpha_{LS}^{(q-1)}}-1)\cdot \mathbf{1}_{(\alpha_{LS}^{(q-1)}<1)})^2]$$

$$= 1+[-\frac{2}{n+2}-\frac{4(n+1)^2}{[n+2]^3(p-k)^2}+\frac{1}{(n+2)^2}$$

$$+\frac{2(n+1)^4}{(n+2)^4(p-q+1)^2}+\frac{4(n+1)^2}{(n+2)^4(p-q+1)^2}]\cdot \mathbf{1}_{(\beta_{LS}^{(k)}<1)}$$

$$\triangleq Q(n,p,q)$$

(38)

From (37) and (38), we can obtain

$$\mathrm{var}\{(1+(\frac{1}{\beta_{LS}^{(q-1)}}-1)\cdot \mathbf{1}_{(\beta_{LS}^{(q-1)}<1)})\}$$

$$= E[(1+(\frac{1}{\beta_{LS}^{(q-1)}}-1)\cdot \mathbf{1}_{(\beta_{LS}^{(q-1)}<1)})^2] \quad . \quad (39)$$

$$- E^2[(1+(\frac{1}{\beta_{LS}^{(q-1)}}-1)\cdot \mathbf{1}_{(\beta_{LS}^{(q-1)}<1)})]$$

$$= Q(n,p,q) - P^2(n,p,q) \triangleq B(n,p,q)$$

Moreover, from [19] (Eq. (43a)), we have

$$\mathrm{var}(\hat{\tau}^{(q-1)}-\sigma^2) = \frac{\sigma^4 \cdot p/n}{(p-q+1)^2} \triangleq \sigma^4 G(n,p,q). \quad (40)$$

Utilizing (37), the mean of $l_q$ in (29) can be derived as

$$E[l_q] = P(n,p,q)(\rho_q \kappa_q + \nu_q - \sigma^2) + \sigma^2. \quad (41)$$

In order to simplify the expression of (41), we let

$$\kappa_q^{LS} = P(n,p,q)\cdot \kappa_q, \quad (42)$$

$$\nu_q^{LS} = P(n,p,q)(\nu_q - \sigma^2) + \sigma^2. \quad (43)$$

Then, (41) can be concisely rewritten as

$$E[l_q] = \rho_q \kappa_q^{LS} + \nu_q^{LS}. \quad (44)$$

From (37), (39) and (40), the variance of $l_q$ in (29) is derived as

$$\mathrm{var}[l_q] = B(n,p,q)\cdot \delta^2(\lambda_q) + \delta^2(\lambda_q)P^2(n,p,q)$$
$$+ B(n,p,q)(\rho_q \kappa_q + \nu_q - \sigma^2)^2 + \sigma^4 G(n,p,q) \quad . (45)$$
$$\triangleq \varsigma_q^2$$

where $\delta(\lambda_q)$ is given in (9). Then, we introduce the following auxiliary statistics:

$$z_q^{LS} = (l_q - \nu_q^{LS})/\kappa_q^{LS} - \sigma^2. \quad (46)$$

According to (7), it is easy to infer that $z_q^{LS}$ follows the following normal distribution:

$$z_q^{LS} \xrightarrow{D} \mathcal{N}\left(\tau_q, (\omega_q^{LS})^2\right). \quad (47)$$

Through simple manipulations, the mean $\tau_q$ and standard deviation $\omega_q^{LS}$ can be easily, derived as:

$$\tau_q = E[z_q] = \frac{E[l_q - \nu_q^{LS}]}{\kappa_q^{LS}} - \sigma^2 = \lambda_q, \quad (48)$$

$$\omega_q^{LS} = \varsigma_q/\kappa_q^{LS}. \quad (49)$$

where $\varsigma_q$ is given in (45). In the next subsection, we will analyze the effect of $\nu_q^{LS}$ in (44) on the estimation performance of the RMT estimator.

*B. Analysis of the effect of the bias term on the estimation performance of the RMT estimator*

Denote by $H_q$ the hypothesis that the true number of sources is $q$. Then, the probability of mis-estimation probability (i.e., estimating the number of signals incorrectly) is defined as

$$P_E = \Pr(\hat{q} \neq q | H_q) = P_{UE} + P_{OE}. \quad (50)$$

where the probability of miss-estimation (under-estimation) $P_{UE}$ and the probability of false alarm (over-estimation) $P_{OE}$ are, respectively, given by

$$P_{UE} = \Pr(\hat{q} < q | H_q), \quad (51)$$

$$P_{OE} = \Pr(\hat{q} > q | H_q). \quad (52)$$

For notational convenience, we denote the right hand side (RHS) of (17) as

$$\varphi_{n,p-q} = \sigma^2 \left(\mu_{n,p-q} + s(\alpha)\sigma_{n,p-q}\right). \quad (53)$$

Then, the condition for the RMT estimator in (17) to determine at least the correct number of signals $q$ is given by

$$l_q > \varphi_{n,p-q}. \quad (54)$$

In order to overcome the noise uncertainty suffered by the

RMT estimator [44], the detection threshold $\varphi_{n,p-q}$ in (54) should be set under the assumption that $l_q$ is arising from noise and thus should be changed as $\varphi_{n,p-q+1}$. Therefore, (54) should be modified as

$$l_q > \varphi_{n,p-q+1}. \quad (55)$$

In order to analyze the effect of the bias term $v_q^{\text{LS}}$ in (44) on the estimation performance of the RMT estimator, we incorporate $v_q^{\text{LS}}$ into the decision condition of the RMT estimator in (55) to introduce an estimator termed as "LS-RMT estimator", which tests the likelihood of the eigenvalue $l_q$ as arising from a signal or from noise according to the following decision condition:

$$l_q - v_q^{\text{LS}} > \varphi_{n,p-q+1}. \quad (56)$$

Then, we derive the increased under-estimation probability of the RMT estimator in (55) incurred by $v_q^{\text{LS}}$ for finite values of $p$ and $n$ as follows.

Firstly, we derive the mis-estimation probability of the LS-RMT estimator in (56). According to (47), the under-estimation probability of the LS-RMT estimator in (56) is given by

$$P_{\text{UE}}^{\text{LS-RMT}}(\omega_q^{\text{LS}}, \kappa_q^{\text{LS}}, \lambda_q, \sigma^2, \varphi_{n,p-q+1}, v_q^{\text{LS}})$$
$$= \Pr\left\{\eta < \frac{[\varphi_{n,p-q+1} + v_q^{\text{LS}}]/\kappa_q^{\text{LS}} - \sigma^2 - \lambda_q}{\omega_q^{\text{LS}}}\right\}. \quad (57)$$

where $\eta \sim \mathcal{N}(0,1)$. Then, the mis-estimation probability of the LS-RMT estimator is given by

$$P_{\text{E}}^{\text{LS-RMT}}(q, v_q^{\text{LS}})$$
$$= \alpha + P_{\text{UE}}^{\text{LS-RMT}}(\omega_q^{\text{LS}}, \kappa_q^{\text{LS}}, \lambda_q, \sigma^2, \varphi_{n,p-q+1}, v_q^{\text{LS}}). \quad (58)$$

Secondly, we derive the mis-estimation probability of the RMT estimator in (55). Without considering $v_q^{\text{LS}}$, the under-estimation probability and over-estimation probability of the RMT estimator are, respectively, given by

$$P_{\text{UE}}^{\text{RMT}}(\omega_q^{\text{LS}}, \kappa_q^{\text{LS}}, \lambda_q, \sigma^2, \varphi_{n,p-q+1})$$
$$= \Pr\left\{\eta < \frac{\varphi_{n,p-q+1}/\kappa_q^{\text{LS}} - \sigma^2 - \lambda_q}{\omega_q^{\text{LS}}}\right\}, \quad (59)$$

$$P_{\text{OE}}^{\text{RMT}}(\sigma_{n,p-q+1}, \sigma^2, v_q^{\text{LS}}) = 1 - F_\beta\left(s(\alpha) - \frac{v_q^{\text{LS}}}{\sigma^2 \cdot \sigma_{n,p-q+1}}\right). \quad (60)$$

From (59) and (60), the mis-estimation probability of the RMT estimator is derived as

$$P_{\text{E}}^{\text{RMT}}(q, v_q^{\text{LS}}) = P_{\text{OE}}^{\text{RMT}}(\sigma_{n,p-q+1}, \sigma^2, v_q^{\text{LS}})$$
$$+ P_{\text{UE}}^{\text{RMT}}(\omega_q^{\text{LS}}, \kappa_q^{\text{LS}}, \lambda_q, \sigma^2, \varphi_{n,p-q+1}). \quad (61)$$

From (58) and (61), the increased under-estimation probability of the RMT estimator incurred by $v_q^{\text{LS}}$ is calculated as

$$\Delta P_{\text{inc-UE}}^{\text{RMT}}(q, v_q^{\text{LS}}) = P_{\text{E}}^{\text{RMT}}(q, v_q^{\text{LS}}) - P_{\text{E}}^{\text{LS-RMT}}(q, v_q^{\text{LS}}). \quad (62)$$

As can be seen from the above analysis, the bias term $v_q^{\text{LS}}$ increases the under-estimation probability of the RMT estimator. Moreover, in the decision condition in (17) of the RMT estimator, the noise is estimated assuming that the eigenvalue being tested is arising from a signal, and thus the noise variance will be under-estimated when the eigenvalue being tested is actually arising from noise. Consequently, the RMT estimator will over-estimate the number of signals in this case. Therefore, it is worthwhile to devise an effective signal number estimation method that can overcome the higher over-estimation probability of the RMT estimator due to the uncertainty in the noise variance estimation and can also overcome the higher under-estimation probability of the RMT estimator incurred by the bias term $v_q^{\text{LS}}$ among eigenvalues for finite system size and finite sample size. In next subsection, we proceed to develop a LS-RMT estimator by utilizing the decision condition in (56).

*C. RMT Estimator via Linear Shrinkage Estimator of Noise Eigenvalues: LS-RMT estimator*

Based on the analysis in Section III. B, we propose an LS-RMT estimator by utilizing the decision condition defined in (56).
. As in (14), the LS-RMT estimator is based on a sequence of hypothesis tests. For $k = 1, 2, \cdots, \min(p,n)-1$, the LS-RMT estimator tests the likelihood of the $k$th eigenvalue $l_k$ as arising from a signal or from noise according to the following two steps.

**Step 1:**

Assuming that $l_{k+1}$, $l_{k+2}$, $\cdots$, $l_p$ is arising from a signal, we obtain the estimates for the noise variance $\sigma_{\text{RMT}}^2(k)$ and the signal eigenvalues $\{\hat{\rho}_i\}_{i=1}^k$ via solving the non-linear equations (15)-(16).

From (11) and (12), we calculate

$$\hat{v}_k = \frac{\hat{\rho}_k}{n}\sum_{i=1}^{k-1}\frac{\hat{\rho}_i}{\hat{\rho}_k - \hat{\rho}_i}, \quad (63)$$

$$\hat{\kappa}_k = 1 + \frac{(p-k)\sigma_{\text{RMT}}^2(k)}{n\hat{\lambda}_k}. \quad (64)$$

Moreover, according to (42) and (43), we calculate

$$\hat{\kappa}_k^{\text{LS}} = P(n,p,k) \cdot \hat{\kappa}_k, \quad (65)$$

$$\hat{v}_k^{\text{LS}} = P(n,p,k)[\hat{v}_k - \sigma_{\text{RMT}}^2(k-1)] + \sigma_{\text{RMT}}^2(k-1). \quad (66)$$

Then, the decision condition for the LS-RMT estimator in (56) is calculated as

$$l_k - \hat{v}_k^{\text{LS}} > \hat{\varphi}_{n,p-k+1}. \quad (67)$$

According to (58), the mis-estimation probability of the LS-RMT estimator is, calculated as

$$P_E^{\text{LS-RMT}}(k, \hat{v}_k^{\text{LS}}) = \alpha + P_{\text{UE}}^{\text{LS-RMT}}(\hat{\omega}_k^{\text{LS}}, \hat{\kappa}_k^{\text{LS}}, \hat{\lambda}_k, \sigma_{\text{RMT}}^2(k), \hat{\varphi}_{n,p-k+1}, \hat{v}_k^{\text{LS}}). \quad (68)$$

**Step 2:**

Then, the proposed LS-RMT estimator tests the likelihood of $l_k$ as arising from a signal or from noise by checking whether the decision condition in (67) is satisfied in the following way: .

(1) If (67) is satisfied, i.e., $l_k - \hat{v}_k^{\text{LS}} > \hat{\varphi}_{n,p-k+1}$, we can infer that $l_k$ is arising from a signal, and $k$ is increased by one. Then, go to Step 1;

(2) Otherwise, i.e., $l_k - \hat{v}_k^{\text{LS}} \leq \hat{\varphi}_{n,p-k+1}$, the number of signals is estimated as $\hat{q} = k-1$.

To summarize, the proposed LS-RMT estimator utilizes the decision condition in (67), and thus can detect the signal eigenvalues immersed in the bias term $\hat{v}_k^{\text{LS}}$. As a result, it overcomes the higher under-estimation probability of the RMT estimator. In addition, the LS-RMT estimator can overcome the noise uncertainty problem suffered by the RMT estimator. Finally, the LS-RMT estimator is summarized in the Algorithm 1.

**Algorithm 1: LS-RMT estimator**

**Input:** Eigenvalues $l_j$ for $j=1,2,\cdots,p$ of $\mathbf{S}_n$, and the significance level $\alpha \in (0,1)$;

For $k=1,2,\cdots,\min(p,n)-1$, test the likelihood of $l_k$ as arising from a signal or from noise as follows:

**Step 1:**

(a) Obtain the estimates for $\{\hat{\rho}_i\}_{i=1}^k$ and $\sigma_{\text{RMT}}^2(k)$ via solving (15)-(16);

(b) Calculate $\hat{\kappa}_k^{\text{LS}}$ in (65), and $\hat{v}_k^{\text{LS}}$ in (66),

**Step 2:**

The LS-RMT estimator tests the likelihood of $l_k$ as arising from a signal or from noise by checking whether the decision condition in (67) in the following way:

**If** ($l_k - \hat{v}_k^{\text{LS}} > \hat{\varphi}_{n,p-k+1}$) **then**

$k \leftarrow k+1$; Go to Step 1.

**Else**

$k \leftarrow k-1$; Go to Step 3.

**End if**

**Step 3:**

Get the signal number estimate: $\hat{q} = k$.

Simulation results are presented to illustrate the performance of the proposed LS-RMT estimator and compare it with the RMT estimator. In the simulations, we utilize a population covariance matrix $\mathbf{\Sigma} = E[\mathbf{x}\mathbf{x}^H]$ that has $q$ signal components with true signal strength $\lambda = [\lambda_1, \lambda_2, \cdots, \lambda_q]$ and Gaussian noise with $\sigma^2 = 1$. The performance measure is the probability of over-estimation $P_{\text{OE}} = \Pr(\hat{q} > q)$, and all results are averaged over 15,000 independent Monte Carlo runs. Fig.1 shows the simulation results for the over-estimation probability $P_{\text{OE}} = \Pr(\hat{q} > q)$ as a function of the system size $p$ with pre-fixed value $\alpha = 0.005$ and $p/n = 1/2$ for *a priori* known noise variance $\sigma^2$: (a) $q=0$; (b) $q=3$ with $\lambda = [150, 120, 100]$, and Fig. 2 shows the corresponding results for the estimated noise variance $\hat{\sigma}^2$.

As can be seen form Fig. 1, the over-estimation probability of the RMT is greater than the pre-fixed value $\alpha = 0.005$ for the case when there is no signal. This is because $v_1^{\text{LS}}$ in (56) is greater than zero, i.e., $v_1^{\text{LS}} > 0$, which increases the over-estimation probability of the RMT. In addition, the over-estimation probability of the RMT is less than the pre-fixed value $\alpha = 0.005$ for the case when there are three signals with $\lambda = [150, 120, 100]$. This is because $v_4^{\text{LS}}$ in (56) is less than zero, i.e., $v_4^{\text{LS}} < 0$, which decreases the over-estimation probability of the RMT. On the contrary, the over-estimation probability of the proposed LS-RMT estimator is around the pre-fixed value $\alpha = 0.005$ for both the case when there is no signal and the case when there are three signals. This is because the LS-RMT estimator utilizes the linear shrinkage (LS) technique to obtain a more accurate estimate for the noise eigenvalue $l_k$ by incorporating $v_k^{\text{LS}}$ into the decision criterion in (56). As a result, the over-estimation probability of the LS-RMT estimator will not be affected by the bias term $v_k^{\text{LS}}$, while the over-estimation probability of the RMT estimator will be affected by this bias term.

As can be seen from Fig. 2, the over-estimation probability of the RMT becomes far greater than the pre-fixed value $\alpha = 0.005$ for the case when the noise level needs to be estimated. This is because the noise variance $\hat{\sigma}_{\text{RMT}}^2(1)$ (or $\hat{\sigma}_{\text{RMT}}^2(4)$) is estimated under the assumption that $l_1$ (or $l_4$) is arising from a signal. Therefore, $\hat{\sigma}_{\text{RMT}}^2(1)$ (or $\hat{\sigma}_{\text{RMT}}^2(4)$) will be smaller than the actual noise variance as $l_1$ (or $l_4$) is actually arising from noise. In turn, the under-estimated noise variance $\hat{\sigma}_{\text{RMT}}^2(1)$ (or $\hat{\sigma}_{\text{RMT}}^2(4)$) will greatly increase the over-estimation probability of the RMT estimator. On the contrary, the over-estimation probability of the proposed LS-RMT estimator is slightly smaller than the pre-fixed value $\alpha = 0.005$ when the system size $p$ (corresponding to the sample size) is relatively small and converges to $\alpha = 0.005$ as the system size $p$ increases for both cases. This is because the noise variance is estimated under the assumption that the eigenvalue being tested is arising from noise in the LS-RMT estimator, and thus will not be under-estimated. As a result, the LS-RMT estimator overcomes the noise uncertainty in the process of noise estimation.

As can be seen from the above discussions, the RMT estimator is affected by the bias term among eigenvalues, and suffers from the uncertainty in the noise variance estimation. In contrast, the proposed LS-RMT estimator overcomes these problems. Therefore, the LS-RMT estimator outperforms the RMT estimator.

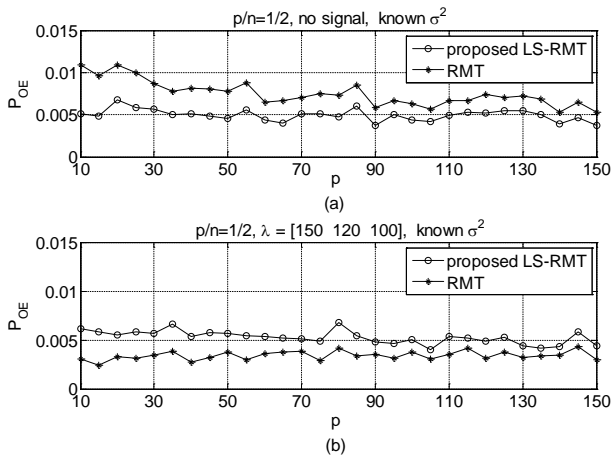

Fig.1. Over-estimation probability $P_{\text{OE}}$ as a function of number of sensors $p$ with pre-fixed value $\alpha = 0.005$ and $p/n = 1/2$ for an *a priori* known noise variance $\sigma^2$: (a) $q = 0$, (b) $q = 3$ with $\lambda = [150, 120, 100]$.

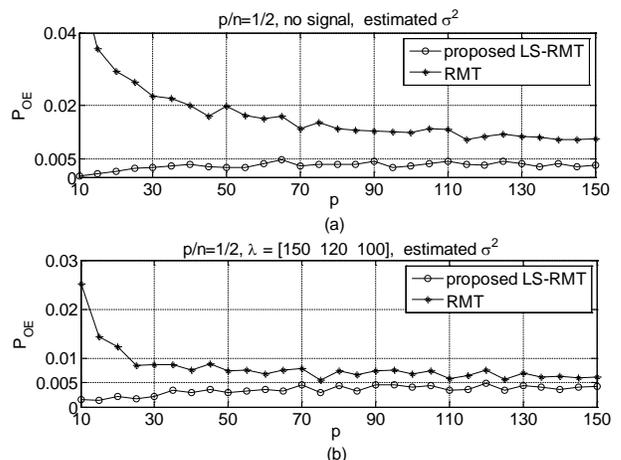

Fig.2. Over-estimation probability $P_{\text{OE}}$ as a function of number of sensors $p$ with pre-fixed value $\alpha = 0.005$ and $p/n = 1/2$ for estimated noise variance $\hat{\sigma}^2$: (a) $q = 0$, (b) $q = 3$ with $\lambda = [150, 120, 100]$.

IV. SIMULATION AND DISCUSSIONS

In this section, we examine the performance of the proposed LS-RMT estimator in Section III.B, and compare it with the existing methods including the classic AIC and MDL estimator [7]-[9], the LS-MDL estimator in [19], the RMT estimator [44], the modified AIC estimator with $C = 2$ in [47], the GE-based estimator in [46], the EV-Differ estimator in [48], and the MDL estimator based on higher-order moments (denoted as HO-MDL estimator) in [50] using Monte Carlo simulations.

In simulations, we assume real valued signals and real valued Gaussian noise unless otherwise stated, the significant level $\alpha$ in both the RMT estimator in (19) and the proposed LS-RMT estimator is set as $\alpha = 0.005$, and we use a population covariance matrix $\Sigma = E[\mathbf{xx}^H]$ that has $q$ unknown signal components with true signal strength $\lambda = [\lambda_1, \lambda_2, \cdots, \lambda_q]$ and Gaussian noise with $\sigma^2 = 1$. For the GE-based estimator in [46], the significance level is set as $\alpha_{\text{GE}} = 0.005$, and the noisy-only sample size is $N = 4n$. For the HO-MDL estimator in [50], the order is set as $r = 4$. The performance measures are the probability of mis-estimation $P_{\text{E}} = \Pr(\hat{q} \neq q)$, the probability of under-estimation $P_{\text{UE}} = \Pr(\hat{q} < q)$, and the probability of the over-estimation $P_{\text{OE}} = \Pr(\hat{q} > q)$. The simulation results are averaged over 15,000 independent Monte Carlo runs.

A. *Performance of the Proposed LS-RMT Estimator for the case $p/n < 1$*

Firstly, we examine the over-estimation probability of the proposed LS-RMT estimator for the case $p/n < 1$. Fig. 3

shows the over-estimation probability of various algorithms as a function of the system size $p$ with fixed ratio $p/n=1/2$ for the case when there is no signal. As can be seen from Fig. 3, the over-estimation probability of the proposed LS-RMT estimator is smaller than the pre-fixed value $\alpha=0.005$ for relatively small system size $p$, and converges to $\alpha=0.005$ as $p$ increases. In contrast, the over-estimation probability of the RMT estimator is much higher than the pre-fixed value $\alpha=0.005$ especially when the system size $p$ is relatively small. This is because the RMT estimator suffers from the uncertainty in noise variance estimation, while the LS-RMT estimator overcomes the noise uncertainty problem by utilizing the linear shrinkage estimator. In addition, the EV-Differ method, the MDL estimator, the modified AIC estimator, and the GE-based estimator have nearly zero over-estimation probability in this case. Finally, the LS-MDL estimator, the HO-MDL estimator, and the conventional AIC estimator have non-negligible over-estimation probability in this case.

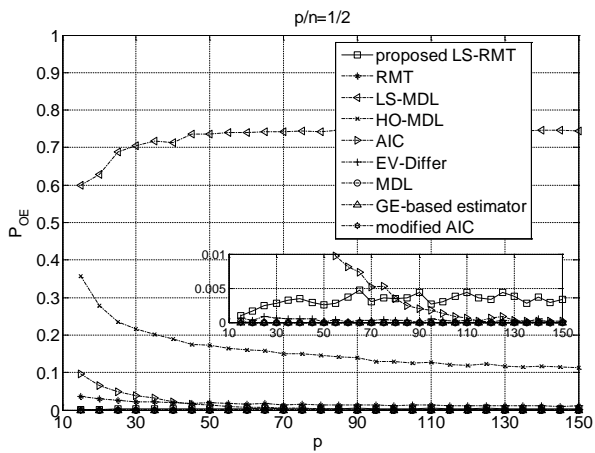

Fig. 3. Comparison of over-estimation probability of various algorithms as a function of system size $p$ with fixed ratio $p/n=1/2$ for the case when there is no signal.

Secondly, we illustrate the estimation performance of the proposed LS-RMT estimator for the case when there are multiple signals. Fig. 4 (a) and (b), respectively, show the mis-estimation probability and over-estimation probability of various algorithms as function of the system size $p$ with fixed ratio $p/n=1/2$ for the case when there are nine signals with $\lambda=[10,10,9,8,7,6,5,4,2.5]$. As can be seen from Fig. 4, the proposed LS-RMT estimator has much better estimation performance than the RMT estimator, the HO-MDL estimator, the LS-MDL estimator, the EV-Differ estimator, the GE-based estimator, the modified AIC estimator, and the MDL estimator. This is because that the LS-RMT estimator considers the bias term among eigenvalues while the other estimators do not consider bias term among eigenvalues. Moreover, it can be seen from Fig. 4(b) that the over-estimation probability of the LS-RMT estimator is smaller than the pre-fixed value $\alpha=0.005$ when the system size $p$ is relatively small, while converges to $\alpha=0.005$ as $p$ increases. In addition, though the over-estimation probability of the LS-MDL estimator, the HO-MDL estimator, the EV-Differ estimator, the MDL estimator, the GE-based estimator, and the modified AIC estimator is zero, the mis-estimation probability of these estimators is far higher than that of the proposed LS-RMT estimator. Finally, though the AIC estimator has better estimation performance than the LS-RMT estimator, it has much larger over-estimation probability than the LS-RMT estimator when the system size $p$ is relatively small, as shown in Fig. 4(b).

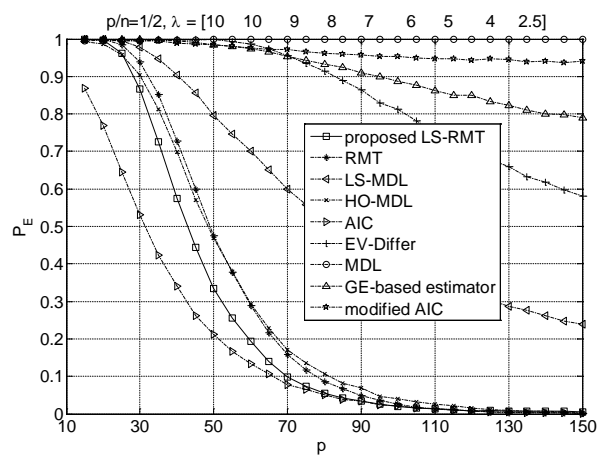

(a) mis-estimation probability

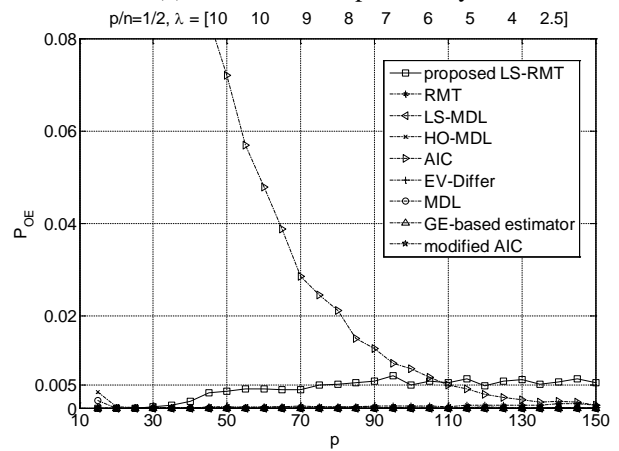

(b) over-estimation probability

Fig. 4. Comparison of (a) mis-estimation probability and (b) over-estimation probability of various algorithms as a function of the system size $p$ with fixed ratio $p/n=1/2$ for the case when there are nine signals with $\lambda=[10,10,9,8,7,6,5,4,2.5]$.

## B. Performance of the proposed LS-RMT estimator for the case $p/n>1$

In this subsection, we illustrate the estimation performance of the proposed LS-RMT estimator for the case when there are multiple signals for the case $p/n > 1$. In this simulation, the simulation results are averaged over 25,000 independent Monte Carlo runs. Fig. 5 (a) and (b), respectively, show the misdetection probability and over-estimation probability of various algorithms as function of the system size $p$ with fixed ratio $p/n = 2$ for the case when there are nine strong signals with $\lambda = [16, 15, 12, 12, 12, 10, 10, 8, 6]$.

As can be seen from Fig. 5, the proposed LS-RMT estimator has much better estimation performance than the RMT estimator, the LS-MDL estimator, the HO-MDL estimator, the EV-Differ estimator, the MDL estimator, the GE-based estimator, and the modified AIC estimator. Moreover, it can be seen from Fig. 5(b) that the over-estimation probability of the LS-RMT estimator is less than the pre-fixed value $\alpha = 0.005$ when the system size $p$ is relatively small and while converges to $\alpha = 0.005$ as $p$ increases. In addition, though the over-estimation probability of the MDL estimator, the AIC estimator, the GE-based estimator and the modified AIC estimator is zero, the mis-estimation probability of these estimators is $100\%$.

when there are nine strong signals with $\lambda = [16, 15, 12, 12, 12, 10, 10, 8, 6]$.

*C. Performance of the proposed LS-RMT estimator for various sample size when the system size is fixed*

In this simulation, the simulation results are averaged over 25,000 independent Monte Carlo runs. Fig. 6 (a) and (b), respectively, show the misdetection probability and the over-estimation probability of various algorithms as a function of sample size $n$ for the case when there are eleven signals with $\lambda = [12, 10, 9, 8, 7, 7, 6, 6, 5, 4, 2.5]$ when the system size $p = 50$.

As can be seen from Fig. 6 (a), the proposed LS-RMT estimator has much better estimation performance than the RMT estimator, the HO-MDL estimator, the LS-MDL estimator, the MDL estimator, the modified AIC estimator, the GE-based estimator, and the EV-Differ method. This is because the LS-RMT estimator can successfully detect the signal eigenvalue immersed in the bias term among eigenvalues. As predicted by the theoretical analysis, the over-estimation probability of the LS-RMT estimator can be controlled around the pre-fixed value $\alpha = 0.005$ for all sample size. In addition, though the over-estimation probability of the LS-MDL estimator, the HO-MDL estimator, the EV-Differ estimator, the MDL estimator, the GE-based estimator, and the modified AIC estimator is zero, the mis-estimation probability of these estimators is far higher than that of the proposed LS-RMT estimator. Finally, though the AIC estimator and have better estimation performance than the LS-RMT estimator for relatively small sample size $n$, it has non-negligible over-estimation probability for all sample size $n$..

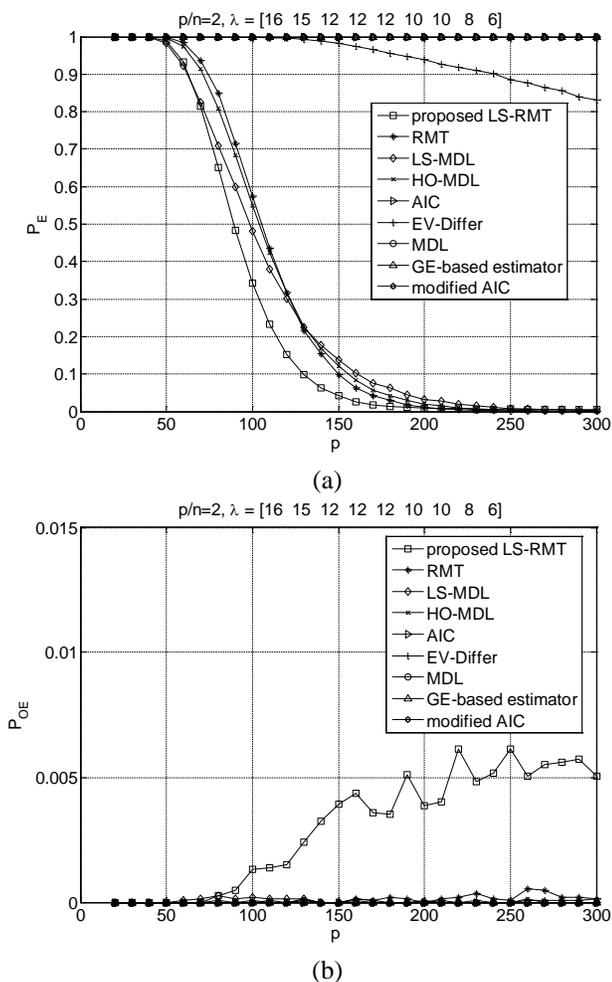

Fig. 5. Comparison of (a) mis-estimation probability and (b) over-estimation probability of various algorithms as a function of the system size $p$ with fixed ratio $p/n = 2$ for the case

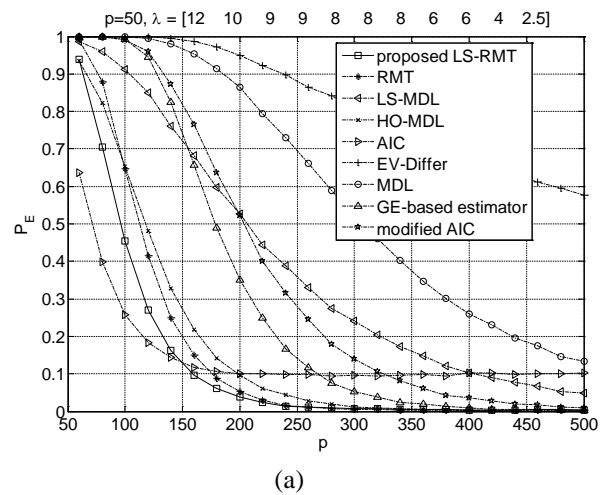

(a)

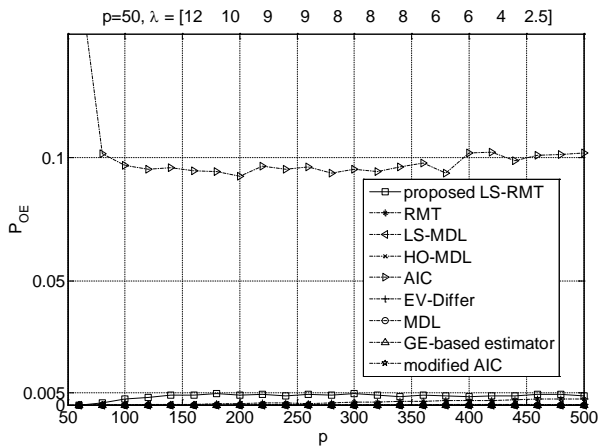

(b)

Fig. 6. Comparison of (a) mis-detection probability and (b) over-estimation probability of various algorithms as a function of sample size $n$ for the case when there are eleven signals with $\lambda = [12, 10, 9, 8, 7, 7, 6, 6, 5, 4, 2.5]$.

### D. On the phase transition phenomenon of the proposed LS-RMT estimator

In [14]-[15], [38]-[41], the authors derived the expressions for phase transition for the signal eigenvalues in the spiked covariance model, which means that the asymptotically detectable signal must have signal strength larger than a threshold $\lambda_{DET} = \sqrt{\gamma}\sigma^2$. In this simulation, we will illustrate the phase transition phenomenon in signal detection of various signal number estimators, and the simulation results are averaged over 25,000 independent Monte Carlo runs.

Fig. 7 (a) and (b), respectively, show the mis-estimation probability and the over-estimation probability of various algorithms as a function of signal strength $\lambda_1$ when $q=1$, $p=160$ and $n=320$. The vertical line in Fig. 7(a) is $\lambda_{DET} = \sqrt{1/2}\sigma^2$. As can be seen Fig. 7, the proposed LS-RMT estimator is significantly better than the HO-MDL estimator, the EV-Differ estimator, the AIC estimator, the MDL estimator, the modified AIC estimator, and GE-based estimator. Though the LS-MDL estimator and the RMT estimator have better estimation performance than the LS-RMT estimator when the signal strength is relatively small, they have higher over-estimation probability than the pre-fixed value $\alpha = 0.005$ as the signal strength $\lambda_1$ becomes large. Moreover, the over-estimation probability of the LS-RMT estimator is less than the he pre-fixed value $\alpha = 0.005$ when the signal strength $\lambda_1$ is relatively small, and converges to $\alpha = 0.005$ as $\lambda_1$ increases.

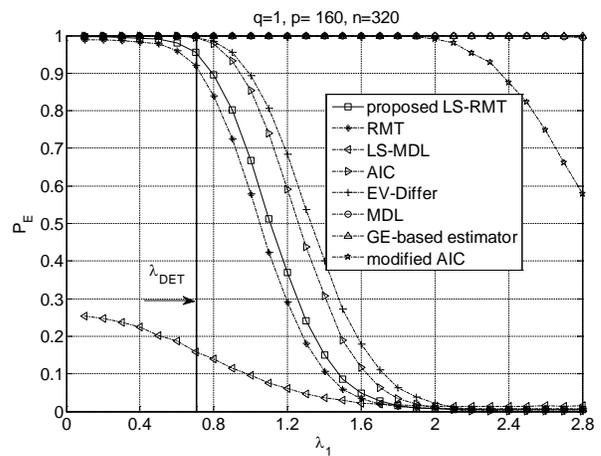

(a) Mis-estimation probability

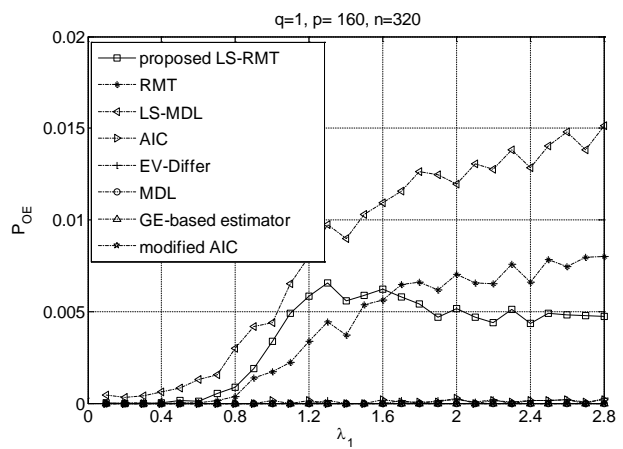

(b) Over-estimation probability

Fig. 7. Comparison of (a) mis-estimation probability and (b) over-estimation probability of various algorithms as a function of signal strength $\lambda_1$ when $q=1$, $p=160$ and $n=320$.

### V. CONCLUSION

As a well-known estimator based on the random matrix theory, the RMT estimator tends to under-estimate the number of signals as signals will be immersed in the non-negligible bias term among eigenvalues for finite sample size. and also suffers from noise uncertainty problem. Moreover, most existing signal number estimation methods do not consider the bias term among eigenvalues, and thus have higher under-estimation probability. In order to overcome the higher down-estimation probability and the noise uncertainty problem of the existing RMT-based estimators, we have proposed a LS-RMT estimator by incorporating the bias term among eigenvalues into the decision criteria of the RMT estimator.

Firstly, we have derived a more accurate expression for the mean (44) and the variance (45) of the noise sample eigenvalue by utilizing the linear shrinkage estimate. Then, we have analyzed the effect of the bias term in (44) among eigenvalues

14on the detection performance of the RMT estimator. Specifically, we have derived the analytical formulas for the increased under-estimation probability of the RMT estimator incurred by the bias term among eigenvalues. Based on these results, the LS-RMT estimator in (56) incorporates the bias term into the decision criterion, and thus can avoid both the higher under-estimation probability and the uncertainty in the noise variance estimation of the RMT estimator. Finally, extensive simulation results show that the proposed LS-RMT estimator has much better detection performance than the existing estimators.